\documentclass{elsart}
\newcommand{\nd}{\noindent}
\newcommand{\nl}{\newline}
\newcommand{\be}{\begin{equation}}
\newcommand{\ee}{\end{equation}}
\newcommand{\ben}{\begin{eqnarray}}
\newcommand{\een}{\end{eqnarray}}

\usepackage{amssymb}

\usepackage[pctex32]{graphics}

\begin{document}

\begin{frontmatter}

\title{Information theory based relations between thermodynamic's 1st. and 2nd. laws }

\author[cbpf]{E. M. F. Curado} and
\ead{evaldo@cbpf.br}\corauth[cor]{Corresponding author.
              Phone / Fax: +55-21-2141-7369}
\author[plata]{A. Plastino}
\ead{plastino@fisica.unlp.edu.ar}

\address[cbpf]{Centro
Brasileiro de Pesquisas Fisicas (CBPF)\\ Rua Xavier Sigaud 150 -
Urca - Rio de Janeiro - Brasil}
\address[plata]{Instituto de F\'{\i}sica (IFLP), Facultad de Ciencias Exactas,\\
                 Universidad Nacional de La Plata and\\
                 Argentina's National Council (CONICET)\\
                 C.C. 727, 1900 La Plata, Argentina}

\begin{abstract}
We focus attention on some particular thermodynamic relations
(PTR). Using information theory concepts we show that, for a
reversible process, microscopic considerations related to these
PTR  make {\it the concomitant informational contents} of the 
first and second laws equivalent. The pertinent demonstration is
obtained when trying to ascertain the corresponding equilibrium
microscopic probability distribution. We also describe other
instances in which the above mentioned informational equivalence
does  not hold.

 \vskip 2mm {\bf  Pacs:\/} 05.70.-a, 05.30.-d, 3.67.-a, 2.50.-r
 \vskip 2mm {\bf  Keywords:\/} Thermodynamics, Microscopic probability distribution, First law,
Second law.

\maketitle

\end{abstract}

\end{frontmatter}

\newpage  \section{Introduction}

 \nd The first  and second laws of thermodynamics are two of
 physics' most important statements.
They constitute strong pillars of our present understanding of
Nature. Of course, statistical mechanics adds an underlying
microscopic substratum  that is able to explain not only these two
laws but the whole of thermodynamics itself
\cite{deslogue,patria,reif,sakurai,katz,Lavenda,PP,1pp,m1}. 
One of its basic
ingredients is an equilibrium microscopic probability distribution
(PD) that controls the population of microstates of the system
under consideration \cite{patria}. We will be mainly concerned
here with changes in the independent external parameters and in
how these changes will affect the microstate-population.

\nd We regard as independent external parameters both extensive
and intensive parameters defining the macroscopic thermodynamic
state of the system. The  extensive parameters, known with
(experimental) certainty, help to define the Hilbert space (HS) in
which the system can be represented. The intensive parameters are
associated with some physical quantities of which  only the
average value is known. They are related to the mean values of the
corresponding operators acting on the HS previously defined. The
eigenvalues of these operators are, therefore, functions of the
extensive parameters defining the HS. The microscopic equilibrium
PD is an {\it explicit} function {\it of the intensive} parameters
and an {\it implicit} function - by means of the eigenvalues of
the above referred to operators (known in average) - {of the
extensive} parameters defining the HS.

\nd In a previous effort \cite{PCPRE} we have shown that
\begin{itemize} \item enforcing the  relation $dU=TdS$ in an infinitesimal
{\it m{\bf i}croscopic} change $p_i \rightarrow p_i+dp_i$ of the
probability distribution (PD) that describes the equilibrium
properties of an arbitrary system \item univocally determines this
PD, and furthermore,
\item the ensuing $\{p_i\}$ coincides with that obtained following
the maximum entropy principle (MaxEnt) tenet of extremizing the
entropy $S$ subject to  an assumedly known  mean value $U$ of the
system's energy. \end{itemize} Such a result undoubtedly exhibits
a {\sf first law-second law relation} ``flavor".   {\it Here} we
wish to further pursue travelling the road paved in \cite{PCPRE}
 by now considering only given infinitesimal {\it m{\bf a}croscopic}
changes (as opposite to the microscopic ones dealt with in
\cite{PCPRE}) in both the (i) intensive and (ii) extensive
parameters of the system, in order to ascertain if such a flavor
becomes more intense and transforms itself into concrete thermal
relations.

\section{Homogenous, isotropic, one-component systems}
\nd Let us start our endeavor by considering  {\it simple},
one-component systems \cite{deslogue}, that is, composed by a
single chemical species, macroscopically homogenous, and isotropic
\cite{deslogue}.
The macroscopic equilibrium thermal state of such a simple,
one-component system is described, in self-explanatory notation,
by $T,\,V,\,N$ \cite{deslogue}. We shall here consider a quite
general information measure $S$ that, according to Kinchin's
axioms for information theory \cite{katz}, depends exclusively on
of the probability distribution $\{p_i\}$ \be \label{1} S=
S(\{p_i\}). \label{here} \ee Contrary-wise, in \cite{PCPRE}, that
uses a different perspective from the present one, a {\it
specific} form for $S$ is used, namely ($f$ is an arbitrary smooth
function of the $p_i$ such that $p f(p)$ is concave)   \be
S=k\sum_{i=1}^W \,p_i\,f(p_i). \label{PRE} \ee We will adopt in
this communication the following notation:  $W$ is the number of
microscopic states, Boltzmann's constant is denoted by $k$, and
the sum runs over a set of quantum numbers, collectively denoted
by $i$ (characterizing levels of energy $\epsilon_i$), that
specify an appropriate basis in Hilbert's space. In \cite{PCPRE}
attention is exclusively focused upon infinitesimal changes in the
$\{p_i\}$ for an $S$ given by (\ref{here}), which is NOT our
perspective here. Let us repeat again that we do not use here
(\ref{PRE}) but rather the more general form (\ref{here}).

\nd Consider further the quantity $U$ that represents the mean
value of the Hamiltonian, and, as befits an homogenous, isotropic,
one-component system in the Helmholtz free energy representation
\cite{deslogue},
\begin{enumerate}
\item as {\it external parameters} the  volume ($V$) and the number of
particles ($N$) (``exactly" known and used to define the Hilbert
space),  \item as {\it intensive variable} the temperature $T$,
associated with the mean value of the internal energy $U$.
\end{enumerate}
\nd The energy eigenvalues of the Hamiltonian $\epsilon_i$ are,
obviously, functions of the volume and of the number of particles,
namely, $\{ \epsilon_i\} = \{ \epsilon_i(V,N) \}$. From now on,
for simplicity, we
take $N$ as  fixed, and   drop thereby the dependence of the
energy eigenvalues on $N$, i.e., $\{ \epsilon_i\} = \{
\epsilon_i(V) \}$. The probability distribution (PD) depends,
then, on the external parameters in the fashion \be \label{probab}
p_i = p_i (T, \epsilon_i(V)). \ee
 Also, let us suppose that $g$ is
an arbitrary  smooth, monotonic function of the $p_i$ such that
$g(0) = 0$ and  $g(1) = 1$. We do not need to require the
condition $\sum_i g(p_i) = 1.$
The mean energy $U$ could be written as 
 \cite{e1,tsallis,Brasil,CT,TMP,e2,NC}, 
\ben \label{4} U = \sum_{i=1}^W \, \,g(p_i) \epsilon_i \, .
\een 
\nd The critical difference between this work and that of
\cite{PCPRE} is to be found in the following assumption, {\sf on
which we entirely base our considerations}: \nl \nd
\fbox{\parbox{5.4in}{the temperature $T$ and the volume $V$
reversibly change in the fashion \be \label{5} T \rightarrow T
\,+\, dT \,\,\, \mbox{and} \,\,\,
V \rightarrow V \,+\, dV. \ee}} \nl \nd  As a consequence of
(\ref{5}), corresponding changes $dp_i$, $dS$, $d\epsilon_i$, and
$dU$ are generated in, respectively, $p_i$, $S$, $\epsilon_i$, and
$U$. Variations in, respectively, $p_i$, $S$, and $U$ write \be
\label{5b} d p_i = \frac{\partial p_i}{\partial T} dT +
\sum_{j=1}^W \frac{\partial p_i}{\partial \epsilon_j}
\frac{\partial \epsilon_j}{\partial V} dV, \ee

\be \label{5c} dS = \sum_{i=1}^W \frac{\partial S}{\partial p_i}
\frac{\partial p_i}{\partial T} dT + \sum_{i, j=1}^W
\frac{\partial S}{\partial p_i} \frac{\partial p_i}{\partial
\epsilon_j} \frac{\partial \epsilon_j}{\partial V} d V, \ee and
\be \label{5d} dU = \sum_{i=1}^W  \frac{\partial g}{\partial p_i}
\frac{\partial p_i}{\partial T} \, \epsilon_i \, dT + \sum_{i,
j=1}^W \frac{\partial g}{\partial p_i} \frac{\partial
p_i}{\partial \epsilon_j} \frac{\partial \epsilon_j}{\partial V}\,
\epsilon_i \, d V
   +  \sum_{i=1}^W g(p_i) \frac{\partial \epsilon_i}{\partial V} d V,
\ee
where, for simplicity, we have considered non-degenerate
levels. Clearly, on account of normalization, the changes in $p_i$
must satisfy the relation

\be \label{norma} \sum_i d p_i =0. \ee

\section{First law
considerations}

The first law of thermodynamics for a reversible process reads
\be
\label{6c}
dU = \delta Q + \delta W = T dS + \delta W,
\ee
where we have used the
Clausius relation $\delta Q = T dS$. Multiplying Eq. (\ref{5c}) by
$T$ we can recast
 Eq. (\ref{6c}) in the fashion \be \label{8b} dU =  T \left( \sum_{i=1}^W
\frac{\partial S}{\partial p_i} \frac{\partial p_i}{\partial T} dT
+ \sum_{i, j=1}^W \frac{\partial S}{\partial p_i} \frac{\partial
p_i}{\partial \epsilon_j} \frac{\partial \epsilon_j}{\partial V} d
V\right) + \delta W . \ee

 \subsection{Changes in the temperature}

\nd Eqs. (\ref{5d}) and (\ref{8b}) must be equal for arbitrary
changes in $T$ and $V$. As these quantities can be changed in an
independent way, let us first consider changes just in $T$.
Enforcing equality in  the coefficients of $dT$ appearing in Eqs.
(\ref{5d}) and (\ref{8b}) we obtain
\be \label{9b} \sum_{i=1}^W
\frac{\partial g}{\partial p_i} \frac{\partial p_i}{\partial T} \,
\epsilon_i \, dT  = T \sum_{i=1}^W \frac{\partial S}{\partial p_i}
\frac{\partial p_i}{\partial T}  dT,
\ee
that must be satisfied
together with [Cf. (\ref{5b})] \be \sum_i\,dp_i=
\sum_i\,\frac{\partial p_i}{\partial T} dT =0. \label{2norma}\ee

We recast now (\ref{9b}) in the fashion
\be \label{10b}
\sum_{i=1}^W \left( \frac{\partial g}{\partial p_i}
\, \epsilon_i - T \frac{\partial S}{\partial p_i}
\right) \frac{\partial p_i}{\partial T} dT \equiv
\,\sum_i\,K_i \frac{\partial p_i}{\partial T} dT =0 .
\ee
Since  the $W$ $p_{i}$'s are not independent
($\sum_{i=1}^W p_i = 1$), we can separate the sum in (\ref{10b})
into two parts, i.e.,
\ben
\label{11c}
 & \sum_{i=1}^{W-1} \left(
\frac{\partial g}{\partial p_i} \, \epsilon_i -
T \frac{\partial S}{\partial p_i}  \right) \frac{\partial
p_i}{\partial T} dT \, + \cr & +
 \left(  \frac{\partial g}{\partial p_W} \, \epsilon_W -
T \frac{\partial S}{\partial p_W}
\right) \frac{\partial p_W}{\partial T} dT = 0.
\een
Picking out level $W$ for special attention
is arbitrary. Any other $i-$level could have been chosen as well,
as the example given below will illustrate.
 Taking into account now that, from Eq.  (\ref{2norma}),
 \be \label{12b}
\frac{\partial p_W}{\partial T} = - \sum_{i=1}^{W-1}
\frac{\partial p_i}{\partial T}, \ee we see that Eq. (\ref{11c})
can be rewritten as \be \label{13b} \sum_{i=1}^{W-1}  \left[
\left( \frac{\partial g}{\partial p_i}  \, \epsilon_i -
T \frac{\partial S}{\partial p_i}    \right)   - \right. \\
\left.   \left(  \frac{\partial g}{\partial p_W}  \, \epsilon_W -
T \frac{\partial S}{\partial p_W}  \right) \right] \frac{\partial
p_i}{\partial T}  dT =0 . \ee As the $W-1$ $p_i$'s  are now
independent, the term into brackets should vanish, which entails
\be \label{14b} \frac{\partial g}{\partial p_i} \epsilon_i - T
\frac{\partial S}{\partial p_i} - \left( \frac{\partial
g}{\partial p_W}  \, \epsilon_W - T \frac{\partial S}{\partial
p_W} \right)   = 0 \, , \ee for all $i =1, \cdots, W-1$. Let us
call the term into parentheses as
\be \label{15b}
K_W =
\frac{\partial g}{\partial p_W}  \, \epsilon_W - T \frac{\partial
S}{\partial p_W} \equiv K.
\ee
We can now cast Eq. (\ref{14b}) as \be
\label{16b} \frac{\partial g}{\partial p_i} \epsilon_i - T
\frac{\partial S}{\partial p_i} - K   = 0; \,\,\,(i =1, \cdots,
W-1), \ee an equation that, hopefully, should yield a definite
expression for $W-1$ of the $p_i$'s, the remaining one being fixed
by normalization.

\vskip 2mm \nd {\bf Example} \vskip 1mm
Consider the Shannon
orthodox instance \ben  S&=&-k\sum_i\,p_i \ln{p_i} \cr
g(p_i)&=&p_i \cr \cr \partial S/\partial p_i
&=&-k[\ln{p_i}+1].\een
\nd Here equation (\ref{16b}) yields \ben
\ln{p_i}&=& -[\frac{\epsilon_i}{kT}+(1-\frac{K}{kT})]; \,\,{\rm i.e.,} \cr
p_i &=&Z^{-1}e^{-\epsilon_i/kT}\cr \ln{Z}&=&1 - K/kT ,\een showing,
as anticipated, that we could have selected any $i-$level other
than $i=W$ above without affecting the final result.

\vskip 4mm \nd In order to better understand the meaning of
equation (\ref{16b}), let us assume now that you wish to {\it
extremize $S$ subject to the constraint of a fixed $U$}, a process
usually referred to as the MaxEnt one \cite{katz}. This is
achieved via a Lagrange multiplier $\beta$. We need also a
normalization Lagrange multiplier $\xi$,

\be  \label{mop}  \delta_{\{\ p_i\}} [S/k - \beta U   - \xi \sum_i\,p_i] =0,   \ee
leading to
\ben  \label{mep1}
 0= \delta_{p_i} \left( S/k
 - \beta  \sum_j  g(p_j)  \epsilon_j -\xi \sum_j p_j\right),
 \een
 implying
 \ben
 \label{17b}
\frac{1}{k}
 \frac{\partial S}{\partial p_i} -
  \beta \frac{\partial g}{\partial p_i} \epsilon_i - \xi = 0 \, ,
 \een
 that, after setting $\xi = - \beta K$, and calling $\beta = 1/kT$, becomes
 \ben
 \label{21b}
T  \frac{\partial S}{\partial p_i} - \frac{\partial g}{\partial p_i} \epsilon_i + K = 0 \, .
\een

\nd Clearly, (\ref{16b}) and (\ref{21b}), are one and the same
equation! Eq. (\ref{21b}) is, in fact, valid for all $W$ states due the
definition of the $i$-independent term $K$ given by Eq. (\ref{15b}).
Therefore, we have
demonstrated that
\begin{itemize}
\item starting from Eq. (\ref{6c}) and \item considering just
changes in the intensive parameter $T$, \item yields the
equilibrium PD obtained using MaxEnt (with the energy constraint).
\end{itemize} It has thus
been shown that, for a simple system, an alternative way to obtain
the equilibrium PD without using MaxEnt exists that only considers
changes in the intensive parameter $T$ of the first law [Eq.
(\ref{6c})].

\subsection{Changes in the extensive  parameter}

\nd Let us now deal with the effect of  changes in the extensive
parameters that  define the Hilbert space in which our system
``lives" and notice that   Eq. (\ref{8b}) can be written in the
fashion \ben \label{8b2simple}
 & dU = \delta Q + \delta W=  TdS  + \delta W =
\cr \cr & T \left(dT \sum_{i=1}^W \frac{\partial S}{\partial p_i}
\frac{\partial p_i}{\partial T} + dV \sum_{i,j=1}^W \frac{\partial
S}{\partial p_i} \frac{\partial p_i}{\partial \epsilon_j}
\frac{\partial \epsilon_j}{\partial V}\right) + \delta W,
\een
i.e.,
\ben \label{entropica} T dS&=& Q_T\,dT +Q_V\,dV;\,\,\,\,\,Q_T=
T \sum_{i=1}^W \frac{\partial S}{\partial p_i} \frac{\partial
p_i}{\partial T}\cr  Q_V&=& T \sum_{i,j=1}^W \frac{\partial
S}{\partial p_i} \frac{\partial p_i}{\partial \epsilon_j}
\frac{\partial \epsilon_j}{\partial V}.\een

Let us now substitute the expression for $(\partial g /
\partial p_i) \epsilon_i$ given by Eqs. (\ref{15b})
and (\ref{16b}),
 \be
\label{16b2}
\frac{\partial g}{\partial p_i} \epsilon_i = T
\frac{\partial S}{\partial p_i} + K;\,\,(i=1,\ldots,W), \ee into
the second term of the R.H.S. of Eq. (\ref{5d}),

\ben \label{2o}
 &  \sum_{i,j =1}^{W} \frac{\partial g}{\partial p_i}
\frac{\partial p_i}{\partial \epsilon_j} \frac{\partial
\epsilon_j}{\partial V}\, \epsilon_i \, d V=
\cr & =
\sum_{i,j=1}^{W} \,[T\frac{\partial S}{\partial
p_i} + K]\frac{\partial p_i}{\partial \epsilon_j} \frac{\partial
\epsilon_j}{\partial V} dV\, = \cr
 &= T\,\sum_{i,j=1}^{W}\,\frac{\partial S}{\partial p_i}\,\frac{\partial
p_i}{\partial \epsilon_j} \frac{\partial \epsilon_j}{\partial
V}dV+ K  \sum_{i,j=1}^{W}
\frac{\partial p_i}{\partial \epsilon_j} \frac{\partial
\epsilon_j}{\partial V} \, dV= \cr
 & = \left( T\,\sum_{i,j=1}^{W}\,\frac{\partial S}{\partial p_i}\,\frac{\partial
p_i}{\partial \epsilon_j} \frac{\partial \epsilon_j}{\partial
V} \right) dV = Q_V\, dV ,
 \een
on account  of the fact that  \be \label{seanula} K
\sum_{i,j=1}^{W} \frac{\partial p_i}{\partial \epsilon_j}
\frac{\partial \epsilon_j}{\partial V} \, dV = 0;\,\,\,{\rm
since}\,\,\, (\partial /\partial V) \sum_i p_i =0.\ee \nd
\fbox{\parbox{5.2in}{We recognize in the term $Q_V \, dV$ of the
last line of (\ref{2o}) a ``volume contribution" to Clausius'
relation $\delta Q=TdS$, an original contribution of the present
work.}} \newpage
  \nd \fbox{\parbox{5.2in}{We see that {\it changes in the equilibrium
PD caused by modifications in the {\sf extensive} parameters
defining the Hilbert space of the system} give also a contribution
to the ``heat part" of the  first law}}. \nl \nd Finally, for Eq.
(\ref{5d}) to become equal to Eq. (\ref{8b2simple}) we have to
demand, in view of the above developments, \ben \label{22b} \delta
W =dV \left[ \sum_i g(p_i) \frac{\partial \epsilon_i}{\partial V}
\right] , \een  the quantity within the brackets being  the mean
value, \ben \label{23b} \left\langle \frac{\partial
\epsilon}{\partial V} \right\rangle = \sum_i g(p_i) \frac{\partial
\epsilon_i}{\partial V}   \, , \een usually associated in the
textbooks with the work done by the system.

\nd Summing up, our analysis of simple systems has shown that
\begin{itemize} \item if we consider the first law written on the form (\ref{6c})
 \item changes in the intensive parameter lead to the obtention of
the equilibrium PD (an alternative way to the MaxEnt principle)
and \item changes in the extensive-Hilbert-space-determining
parameter lead to two contributions \begin{enumerate} \item one
related to heat and \item the other related to work.
\end{enumerate} \end{itemize}

\section{More general systems}

\nd More general systems can be considered by: (i) considering
multi-component ones, and/or (ii) by adding additional extensive
variables \cite{deslogue}. As the numbers of components
$N_1,\cdots,N_S$ of a multi-component system are themselves
extensive variables, we can considerably amplify the preceding
considerations by considering, instead of just one extensive
quantity (volume) as we did before, $M$ of them $X_1,\cdots,X_M$
\cite{deslogue}. The conclusions reached in the previous Section
can then be straightforwardly generalized. One would deal  with
one  intensive parameter ($T$) and  $M$
extensive-Hilbert-space-determining ones, where one of them would
be the volume. Eq. (\ref{entropica}) generalizes to

\ben \label{2entropica} T dS&=& Q_T\,dT +
\sum_{\nu=1}^M\,Q_\nu\,dX_\nu;\,\,\,\,\,Q_T= T \sum_{i=1}^W
\frac{\partial S}{\partial p_i} \frac{\partial p_i}{\partial T}\cr
Q_\nu&=&T \sum_{i,j=1}^W \frac{\partial S}{\partial p_i}
\frac{\partial p_i}{\partial \epsilon_j} \frac{\partial
\epsilon_j}{\partial X_\nu},\een
while $dU$ is now

\ben \label{5ddos} & dU = \sum_{i=1}^W  \frac{\partial g}{\partial
p_i} \frac{\partial p_i}{\partial T} \, \epsilon_i \, dT + \cr & +
\sum_{\nu=1}^M\,\left[\sum_{i, j=1}^W \frac{\partial g}{\partial
p_i} \frac{\partial p_i}{\partial \epsilon_j} \frac{\partial
\epsilon_j}{\partial X_\nu}\, \epsilon_i \, d X_\nu
   +  \sum_{i=1}^W g(p_i) \frac{\partial \epsilon_i}{\partial X_\nu} d X_\nu \right].
\een where, again for simplicity we have considered non-degenerate
levels. The changes in $p_i$ must of course still  satisfy the
relation

\be \label{normados} \sum_i d p_i =0. \ee
 Eq. (\ref{5ddos}) gets simplified, following the lines developed
 above, to

\ben \label{new8b2simple} & dU = \delta Q + \delta W=  TdS  +
\delta W= \cr \cr & T \left(dT \sum_{i=1}^W \frac{\partial
S}{\partial p_i} \frac{\partial p_i}{\partial T} +
\sum_{\nu=1}^M\,dX_\nu \left[ \sum_{i,j=1}^W \frac{\partial
S}{\partial p_i} \frac{\partial p_i}{\partial \epsilon_j}
\frac{\partial \epsilon_j}{\partial X_\nu}\right]\right) + \delta
W, \een
with
\be
\label{trabajo}
 \delta W= \sum_{\nu=1}^M\,\delta W_\nu,
\ee
where

\ben \label{dos22b} \delta W_\nu =dX_\nu \left[ \sum_{i=1}^W
g(p_i) \frac{\partial \epsilon_i}{\partial X_\nu} \right] , \een
involving the mean value \ben \label{dos23b} \left\langle
\frac{\partial \epsilon}{\partial X_\nu} \right\rangle = \sum_i
g(p_i) \frac{\partial \epsilon_i}{\partial X_\nu}   \, , \een
usually associated in the textbooks with the work done by the
system in changing the external parameter $X_\nu$.

\section{Carnot cycle revisited}

\nd Let us now consider an ideal gas performing a Carnot cycle,
constituted by two isothermal and two adiabatic ``trajectories".
The external parameters that are changing in this reversible
transformation are: (i) an intensive parameter, the temperature
($T$), and (ii) an extensive parameter, the volume ($V$).

\subsection{Isothermal reversible transformation}

In an isothermal transformation $dT = 0$, implying $dU = 0$. Eqs.
(\ref{5c}) and (\ref{5d})  write now

\be
\label{40} dS = \sum_{i, j}
\frac{\partial S}{\partial p_i} \frac{\partial p_i}{\partial
\epsilon_j} \frac{\partial \epsilon_j}{\partial V} d V ,
\ee
and
\be
\label{41} dU = \sum_{i,
j} \frac{\partial g}{\partial p_i} \frac{\partial
p_i}{\partial \epsilon_j} \frac{\partial \epsilon_j}{\partial V}\,
\epsilon_i \, d V
   +  \sum_{i=1}^W g(p_i) \frac{\partial \epsilon_i}{\partial V} d V \, .
\ee Clearly, from Eq. (\ref{40}), the heat term in this
transformation can be identified with \be \label{42} \delta Q = T
dS = T \sum_{i, j} \frac{\partial S}{\partial p_i} \frac{\partial
p_i}{\partial \epsilon_j} \frac{\partial \epsilon_j}{\partial V} d
V , \ee and has its origin in the way changes in the energy
levels, due to the volume transformation, affect the population of
the microscopic states.  The ensuing PD-changes have an impact on
the entropy, since heat is  exchanged in this transformation. As
we have shown in Eq. (\ref{2o}) \be \label{43} \delta Q = T
\sum_{i, j} \frac{\partial S}{\partial p_i} \frac{\partial
p_i}{\partial \epsilon_j} \frac{\partial \epsilon_j}{\partial V} d
V = \sum_{i, j} \frac{\partial g}{\partial p_i} \frac{\partial
p_i}{\partial \epsilon_j} \frac{\partial \epsilon_j}{\partial V}\,
\epsilon_i \, d V \ee and, as $dU = 0$ from Eq. (\ref{41}), we see
that the heat term should be equal to minus the work term,
implying \be \label{44}  T \sum_{i, j} \frac{\partial S}{\partial
p_i} \frac{\partial p_i}{\partial \epsilon_j} \frac{\partial
\epsilon_j}{\partial V} d V = \sum_{i, j} \frac{\partial
g}{\partial p_i} \frac{\partial p_i}{\partial \epsilon_j}
\frac{\partial \epsilon_j}{\partial V}\, \epsilon_i \, d V =
   -  \sum_{i=1}^W g(p_i) \frac{\partial \epsilon_i}{\partial V} d V \, ,
\ee where the work done is given by the term \be \label{45}
\sum_{i=1}^W g(p_i) \frac{\partial \epsilon_i}{\partial V} d V =
\left\langle \frac{\partial \epsilon}{\partial V} \right\rangle =
\delta W \, , \ee as it is usually understood.  Eq. (\ref{43}) is
 a microscopic recipe for evaluating  the heat term in an isothermal
reversible transformation {\it without computing first the work
term}.

\subsection{Adiabatic transformation}

\nd Since here $dS = 0$,  Eq. (\ref{5c}) yields \be \label{46}
\sum_{i} \frac{\partial S}{\partial p_i} \frac{\partial
p_i}{\partial T} dT = - \sum_{i, j} \frac{\partial S}{\partial
p_i} \frac{\partial p_i}{\partial \epsilon_j} \frac{\partial
\epsilon_j}{\partial V} d V \, , \ee and the equation for the
associated trajectory  in the ($T,V$)-plane of the adiabatic
reversible transformation can be given as \be \label{47}
\frac{dT}{dV}  =  - \left( \sum_{i, j} \frac{\partial S}{\partial
p_i} \frac{\partial p_i}{\partial \epsilon_j} \frac{\partial
\epsilon_j}{\partial V} \right) / \left(  \sum_{i} \frac{\partial
S}{\partial p_i} \frac{\partial p_i}{\partial T} \right)   \, .
\ee Using Eqs. (\ref{9b}), (\ref{2o}) and (\ref{46}) we can see
that Eq. (\ref{5d}) can be written, for this transformation, as
\be \label{48} dU = \sum_i g(p_j) \frac{\partial
\epsilon_j}{\partial V} dV \, , \ee and it contains only the work
term, as it is well-known.

\subsection{Process without work}

\nd In this kind of process we should have \be \label{49} dU =
\delta Q = T dS \, , \ee that are due, according to the present
viewpoint,  to changes affecting only  the  intensive parameter
$T$. If there is no change in any of the extensive parameters
defining the Hilbert space, there cannot  be a work term, as the
eigenvalues of the operators in this Hilbert space will not
change. Thus, changes in the intensive parameters induce
 only  heat-changes, i.e., $dS-$variations, and not $W-$ones.

\section{Conclusions}

\nd This paper advanced a new microscopic picture of the ``first
law-second law" marriage for any arbitrary entropic functional of
the underlying probability distribution. We have
\begin{itemize}
\item exhibited the microscopic details of the way in which
\begin{enumerate}
\item  changes in the macroscopic intensive variables affect only
the heat part of the first law,  while \item displaying the same
informational content that we can be obtained using
MaxEnt.\end{enumerate} \item Changes in the macroscopic extensive
variables used to define the Hilbert space of system affect both
the heat and the work part of the first law. Recourse to first-law
considerations allows one to visualize the way in which {\sf
changes in the extensive} variables affect the PD and, in turn,
{\it do give a contribution to heat terms}, a new result.
\end{itemize}

\noindent
{\bf Acknowledgments:} The authors thank Renio S. Mendes for 
valuable discussions. E.M.F.C. also thanks PRONEX and CNPq (Brazil) 
for partial support.

  \end{document}